\documentclass[aps,prd,amsmath,onecolumn,notitlepage,showpacs,superscriptaddress,nofootinbib,usenatbib]{revtex4-1}

\usepackage[utf8x]{inputenc}
\usepackage[T1]{fontenc}
\usepackage{color}
\usepackage{dcolumn}
\usepackage{bm}

\usepackage{amsmath}
\usepackage{graphicx}
\usepackage[colorinlistoftodos]{todonotes}
\usepackage[colorlinks=true, allcolors=blue]{hyperref}
\newcommand{\mat}[1]{\mbox{\boldmath{$#1$}}} 
\newcommand{\BibitemShut}[1]{}

\begin{document}

\title{Isotropy of low redshift type Ia Supernovae: A Bayesian analysis}

\author{U. Andrade}
\email{uendertandrade@on.br}
\affiliation{Observat\'orio Nacional, 20921-400, Rio de Janeiro, RJ, Brasil}

\author{C. A. P. Bengaly}
\email{carlosap87@gmail.com}
\affiliation{Department of Physics \& Astronomy, University of the Western Cape, 7535, Cape Town, South Africa}

\author{J. S. Alcaniz}
\email{alcaniz@on.br}
\affiliation{Observat\'orio Nacional, 20921-400, Rio de Janeiro, RJ, Brasil}
\affiliation{Physics Department, McGill University, Montreal QC, H3A 2T8, Canada}

\author{B. Santos}
\email{thoven@on.br}
\affiliation{Observat\'orio Nacional, 20921-400, Rio de Janeiro, RJ, Brasil}

\begin{abstract}
The standard cosmology strongly relies upon the Cosmological Principle, which consists on the hypotheses of large scale isotropy and homogeneity of the Universe. Testing these assumptions is, therefore, crucial to determining if there are deviations from the standard cosmological paradigm. In this paper, we use the latest type Ia supernova compilations, namely JLA and Union2.1 to test the cosmological isotropy at low redshift ranges ($z<0.1$). This is performed through a Bayesian selection analysis, in which we compare the standard, isotropic model, with another one including a dipole correction due to peculiar velocities.  The full covariance matrix of SN distance uncertainties are taken into account. We find that the JLA sample favors the standard model, whilst the Union2.1 results are inconclusive, yet the constraints from both compilations are in agreement with previous analyses. We conclude that there is no evidence for a dipole anisotropy from nearby supernova compilations, albeit this test should be greatly improved with the much-improved datasets from upcoming cosmological surveys.          
\end{abstract}

\maketitle

\section{Introduction}

One of the cornerstone of modern cosmology is the so-called Cosmological Principle (CP), which consists of a statement that the Universe is statistically isotropic and homogeneous on large scales. A mathematical representation of these symmetries is given by the Friedmann-Lama\^itre-Robertson-Walker (FLRW) metric, and based on it one may build up homogeneous and isotropic cosmological models such as the standard $\Lambda$CDM cosmology. This scenario seems to best account for our current observations, namely, the angular power spectrum of the cosmic microwave background (CMB) temperature fluctuations~\cite{ade2016planck1}, the clustering of matter which describes the large-scale structure of the Universe~\cite{alam2017clustering, ata2017clustering}, besides the cosmic distances from Type Ia Supernovae (SNe)~\cite{suzuki2011hubble, betoule2014improved}, and age measurements of passively evolving galaxies over the cosmic evolution~\cite{Alcaniz:1999kr,Alcaniz:2001uy,moresco20166}. 

Although the validity of $\Lambda$CDM model has been put under scrutiny along the last decades, some of its underlying hypotheses, such as the CP, have always been assumed to hold true in most of the analyses. Since it comprises one of the most fundamental hypothesis of the standard cosmological model, it is crucial to directly assess its validity against cosmological observations, as significant departures from statistical isotropy and homogeneity would require a complete reformulation of the standard cosmological paradigm\footnote{One of the most common approaches to test cosmological isotropy in the literature consists on probing consistency between the dipole signal seen in the CMB global temperature, which is ascribed to Doppler and aberration effects due to our relative motion~\cite{kogut1993dipole, aghanim2014planck, hoffman2017dipole}. We will not pursue this approach here, but rather look for the statistical validity of possible anisotropic signatures in the low-redshift Universe.}. 

In the recent years, some authors reported potential deviations from the CP, e.g., the presence of an unexpected large velocity flow of galaxy clusters via kinematic Sunyaev-Zeldovich  effect (kSz)~\cite{kashlinsky2009measurement, kashlinsky2010new, kashlinsky2011measuring, atrio2015probing}, large-angle features in the CMB temperature map~\cite{schwarz2015cmb, ade2016planck2}, as well as a large dipole anisotropy in the counts of radio sources~\cite{singal2011large, rubart2013cosmic, colin2017high, bengaly2017probing}. However, others authors claim no significant evidence for some of these signals, at least with the present uncertainties and systematics. For instance,~\cite{feindt2013measuring} ruled out, at $\sim 4 \sigma$ confidence level, a bulk flow of $\sim 1000 \; \mathrm{km/s}$ as seen by \cite{kashlinsky2010new} using low redshift SN compilations. A similar result was confirmed by~\cite{mathews2016detectability}, also using SNe, and by~\cite{ade2014planck} through Planck's first kSZ data release. On the other hand, the CMB features and the large anisotropy on the radio sky remain open issues. 

Therefore, we need to clarify whether some of these puzzles arise due to the limitations of current cosmological datasets, or whether they are actual indications of physics beyond the standard model. In order to do this, we use the most recent compilations of SNe, namely, the Joint Light curve Analysis (JLA) \cite{betoule2014improved} and Union2.1 \cite{suzuki2011hubble}, to test the local isotropy ($0.015 < z < 0.1$) in the Universe, similarly to previous works~\cite{gordon2008determining, colin2011probing, dai2011measuring, ma2011peculiar, rathaus2013studying, feindt2013measuring, appleby2015probing}. The novelty of this work is the Bayesian selection analysis performed in the low-$z$ SN data, so we can compare how does a dipole modulation in the luminosity distance due to local peculiar velocities, as proposed by~\cite{bonvin2006dipole}, can characterize the observational data with respect the standard cosmographic analysis\footnote{Even though the luminosity distance modification presented in~\cite{bonvin2006dipole} consists of a correction rather than a different model, it provides us a framework to estimate the constraints on the velocity dipole, and thus the validity of the isotropy hypothesis in the local Universe.}. If we find strong evidence for a higher velocity dipole than obtained in previous analyses, this would hint at a possible breakdown of the FLRW description at the $z<0.1$ scales. 

The paper is organised as follows: \S{2} presents the models considered in this analysis; \S{3} is dedicated to the description of the observational samples;  \S{4} presents the Bayesian selection method adopted throughout our analysis; \S{5} discusses our results; the concluding remarks are discussed in \S{6}.


\section{Models}

\subsection{Cosmographic model}

Assuming that the FLRW metric holds true, we can expand the scale factor around the present time, and then measure distances regardless the dynamics of the Universe. This is the well-known cosmographic approach~\cite{Weinberg:1972kfs,visser2004jerk,Lazkoz:2013ija}~\footnote{We caution that this approach suffers from some caveats, specially in high-redshift ranges, as pointed out in~\cite{busti2015cosmography}. These problems do not arise at $z < 0.1$, though.}. The luminosity distance reads
\begin{eqnarray}\label{background}
D_{L}(z) = \frac{c}{H_{0}} \left[z + (1 - q_0)\frac{z^2}{2} + O(z^3)\right] \;,
\end{eqnarray}
where $z$ is the redshift observed in the comoving rest frame with respect to the expansion of the Universe, $H_0$ and $q_0$ are the Hubble constant and decelerating parameter at present time, respectively, being $c$ the speed of light given in {\rm{km/s}}. Cosmography is well-known to suffer from divergences in the Taylor Expansion. Finite truncations could give rise to systematics. Thus, in order to avoid this issue we parameterise the redshift variable by $y \equiv \frac{z}{1 + z}$~\cite{Cattoen:2007id, Vitagliano:2009et}. Therefore,  rewriting the luminosity distance in terms of  $y$ , usually referred to as \textit{y-redshift}, we have,
\begin{eqnarray}\label{background y-redshift}
D_{L} (y)  = \frac{c}{H_0} \left[y + (3 - q_0)\frac{y^2}{2} + O(z^3)\right]  \;.
\end{eqnarray}

 This expression is related to Eq.~(\ref{eq:dist_mod1}) through
\begin{eqnarray}\label{eq:dist_mod2}
\mu = 5\log_{10}\left({D_{L}(y) \over \rm{Mpc}}\right) + 25 \;.
\end{eqnarray}
As shown in the Eq. (\ref{background y-redshift}), the luminosity distance depends only on $H_0$ and $q_0$ up to the second order in redshift. Therefore, we restrict our analysis up to that order. It is worth mentioning that we estimated the error on $D_{L}$ that this truncation may result and found that the third order term in Eq. (\ref{background y-redshift}) differs in less than $1\%$ at $z = 0.1$ compared to the second order term in the same expression. Hereafter we will refer to Eq. (\ref{background y-redshift}) as the \emph{reference model}. 

\subsection{Dipole modulation}

As pointed out in~\cite{bonvin2006fluctuations}, the directional luminosity distance can be expanded in spherical harmonics leading to a observable multipoles, $C_{\ell}(z)$. The dipole term, $C_{1}(z)$, has a minor contribution from lensing effect, whereas our peculiar motion has a huge impact on it. Neglecting higher order terms than the dipole, one can write the directional luminosity distance as following:
\begin{eqnarray}
D_{L}(z, \textbf{n}) = D_{L}^{(0)}(z) + D_{L}^{(1)}(z)(\textbf{n} \cdot \textbf{e}) \;,
\end{eqnarray}
where $D_{L}^{(0)}(z)$ is set up to be the reference model as in Eq.~(\ref{background y-redshift}), and $D_{L}^{(1)}(z)(\textbf{n} \cdot \textbf{e})$ is obtained in a perturbed Friedmann Universe, i.e.,
\begin{eqnarray}\label{dipole}
D_{L}^{(1)}(z) (\textbf{n} \cdot \textbf{e}) = \frac{(1 + z)^2}{H(z)} (\mathbf{n} \cdot \mathbf{v}_0) \;,
\end{eqnarray}
being $\textbf{v}_0$ the peculiar velocity, $H(z)$ the Hubble parameter, and $\mathbf{n}$ and $\mathbf{e}$ correspond to unit vectors denoting the SN position in the sky and the bulk flow direction, respectively. Thus, the luminosity distance can be rewritten such as
\begin{eqnarray}\label{CD}
& & D_{L} = \frac{c}{H_0} \left[y + (3 - q_0)\frac{y^2}{2} \right] + \frac{v_{bf}\cos{\phi}}{H(y)}\left(z/y \right)^2 \;, 
\end{eqnarray}
where $v_{bf}$ is the bulk flow velocity, and $\cos{\phi} = (\mathbf{n} \cdot \mathbf{v}_{bf})$ denotes the angle between each SN location and the bulk flow velocity direction. We will henceforth refer to this model, where $D_{L}$ is corrected by the bulk flow velocity at first order, as the {\emph{Dipole model}}.


\section{The observational datasets} 
\label{sec:2}

\begin{figure*}
\includegraphics[width=0.40\textwidth, height=5.0cm]{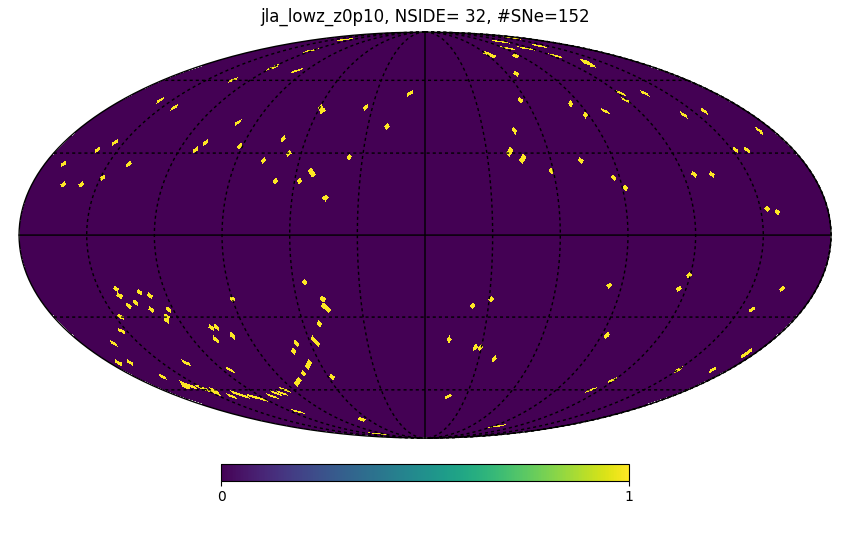}
\includegraphics[width=0.40\textwidth, height=5.0cm]{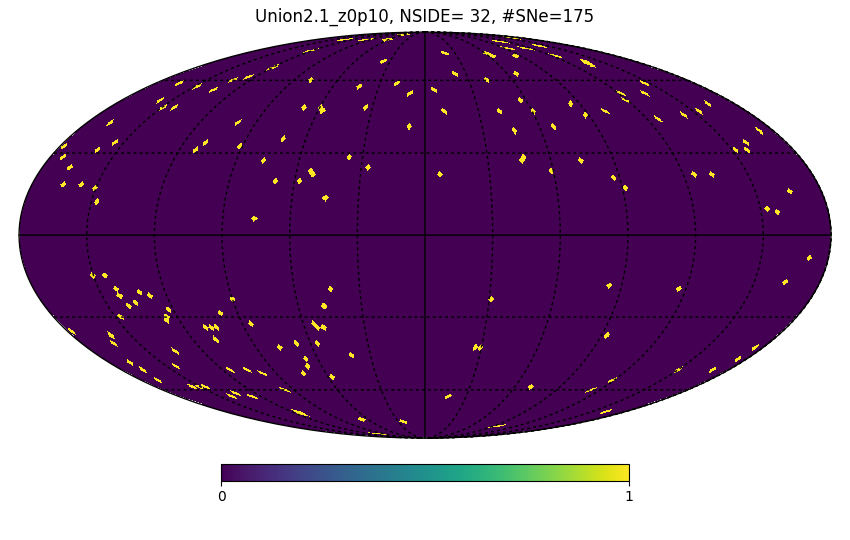}
\caption{Mollweide projections of the SN celestial distribution for JLA (left panel) and Union2.1 (right panel) datasets.}
\label{fig:sn_maps}
\end{figure*}

Our analysis is performed using both the JLA and the Union2.1 samples, as obtained from both The Paris Supernova Cosmology Group and The Supernova Cosmology Project websites\footnote{\href{http://supernovae.in2p3.fr/sdss_snls_jla/ReadMe.html}{http://supernovae.in2p3.fr/}}\footnote{\href{http://supernova.lbl.gov/union/}{http://supernova.lbl.gov/}}, respectively. The JLA sample consists of a set of 740 spectroscopically confirmed SNe, while Union2.1 compilation encompasses 580 objects. 

The modulus distance of each SNe is given by
\begin{eqnarray}\label{eq:dist_mod1}
\mu = m_B - (M_B - \alpha \times x_1 + \beta \times c_l)\;,
\end{eqnarray}
where $m_B$ is the observed peak magnitude in the rest frame $B$ band, $\alpha, \beta, M_B$ are nuisance parameters in the distance modulus estimate that account, respectively, for corrections on the shape ($x_1$) and color ($c_l$) of the SN light-curves, and absolute magnitude in the same $B$ band. In addition, $d_L$ corresponds to the dimensionless luminosity distance, i.e., $d_L = (H_0/c)D_L$, whereas $\mathcal{M}$ is given by

\begin{equation}
\label{eq:mathcal}
\mathcal{M} \equiv 25 + \log_{10}{\left(\frac{c}{H_0 \times 1Mpc}\right)} + M_B\;. 
\end{equation}
Since $H_0$ is degenerated with the absolute magnitude $M_B$, we thus treat it as a nuisance parameter along with $\alpha$ and $\beta$ in our analyses.

After selecting only the objects lying in the $z<0.1$ range in both compilations, we plot their distribution in the sky as displayed in Fig.~\ref{fig:sn_maps}. These maps were constructed using the {\sc HEALPix} software package~\cite{gorski2005healpix}.



\begin{table}[h!]
	\caption{The flat priors $\mathcal{U}(a, b)$ assumed throughout our analysis.}
	\renewcommand{\arraystretch}{1.5}
	\centering
	\medskip
	\label{priors}
	\begin{tabular}{ccc}
		\hline
		\hline
		Parameters & Model associated & Prior  \\ 
		\hline
		$q_0$ & All & $\mathcal{U}(-0.67,-0.49)$  \\ 
		$v_{bf}$ & $DM$ & $\mathcal{U}(0, 500)$  \\ 
		$l$ & $DM$ & $\mathcal{U}(0, 360)$  \\ 
		$b$ & $DM$ & $\mathcal{U}(-90, 90)$ \\
		\hline
	\end{tabular} 
\end{table}


\section{Bayesian Analysis}

According to the Bayes' theorem, there is a relationship between a given model and a set of data and the information one might know beforehand. Mathematically, the Bayes' theorem reads
\begin{eqnarray}\label{Bayes Theorem}
P(\Theta| D, M) = \frac{\mathcal{L}(D|\Theta, M) \mathcal{P}(\Theta, M)}{\mathcal{E}(D, M)} \;,
\end{eqnarray}
where $P$ is the posterior, i.e., the probability that a set of parameters ($\Theta$) arise from the data given a model $M$; $\mathcal{L}$ is the likelihood, thus describing how we believe the data are distributed; $\mathcal{P}$ is the prior knowledge about the parameters; $\mathcal{E}$ is the evidence that the data were drawn from a given model. 

In model fitting the denominator of Eq. (\ref{Bayes Theorem}) can be ignored, as it only consists of a normalization constant. Thus, we can write a posterior marginalized over the set of parameter we are not interested in ( the so-called nuisance parameters of \S{2}), such as
\begin{eqnarray}
P(\theta| D, M) \propto \int \mathcal{L}(D|\Theta, M) \mathcal{P}(\Theta, M) \, d\phi \;.
\end{eqnarray}
On the other hand, the denominator in Eq. (\ref{Bayes Theorem}) plays the rule in model section, since it tells us how likely the data were sample from the model. Then, we can use it to compare between different models, so that the evidence can be computed from the following integral over the full parameter space of the underlying model: 
\begin{eqnarray}
\mathcal{E}(D, M) = \int_{M} \mathcal{L}(D|\Theta, M) \mathcal{P}(\Theta, M) d\Theta \;.
\end{eqnarray}

In order to perform such comparison between models we compute the ratio,
\begin{eqnarray}\label{ratio}
B_{12} \equiv \frac{\mathcal{E}_1}{\mathcal{E}_2} \;,
\end{eqnarray}
which is usually referred to as Bayes' Factor, and we use it to calculate the Jeffreys scale \cite{jeffreys1961theory} in order to discriminate two competing models. Here, we use a modified version of Jeffreys scale, as suggested by \cite{trotta2008bayes}, where the strength of the evidence is regarded as \text{inconclusive} when $|\ln{B_{12}}| < 1$, \text{weak} in the cases where $|\ln{B_{12}}| > 1$, \text{moderate} if $|\ln{B_{12}}| > 2.5$, and \text{strong} for cases where  $|\ln{B_{12}}| > 5$. It is worthy noting that $\ln{B_{12}} > 1$ indicates that model one, $M_1$, is favoured in comparison to model two, $M_2$. Otherwise, $\ln{B_{12}} < - 1$ would support model $M_2$ (for some recent application of Bayesian model selection in cosmology, see~\cite{Santos:2016sti,Santos:2017alg, Dhawan2017} and references therein).   

\begin{figure}[!t]
\includegraphics[width=4.5cm, height=4.5cm]{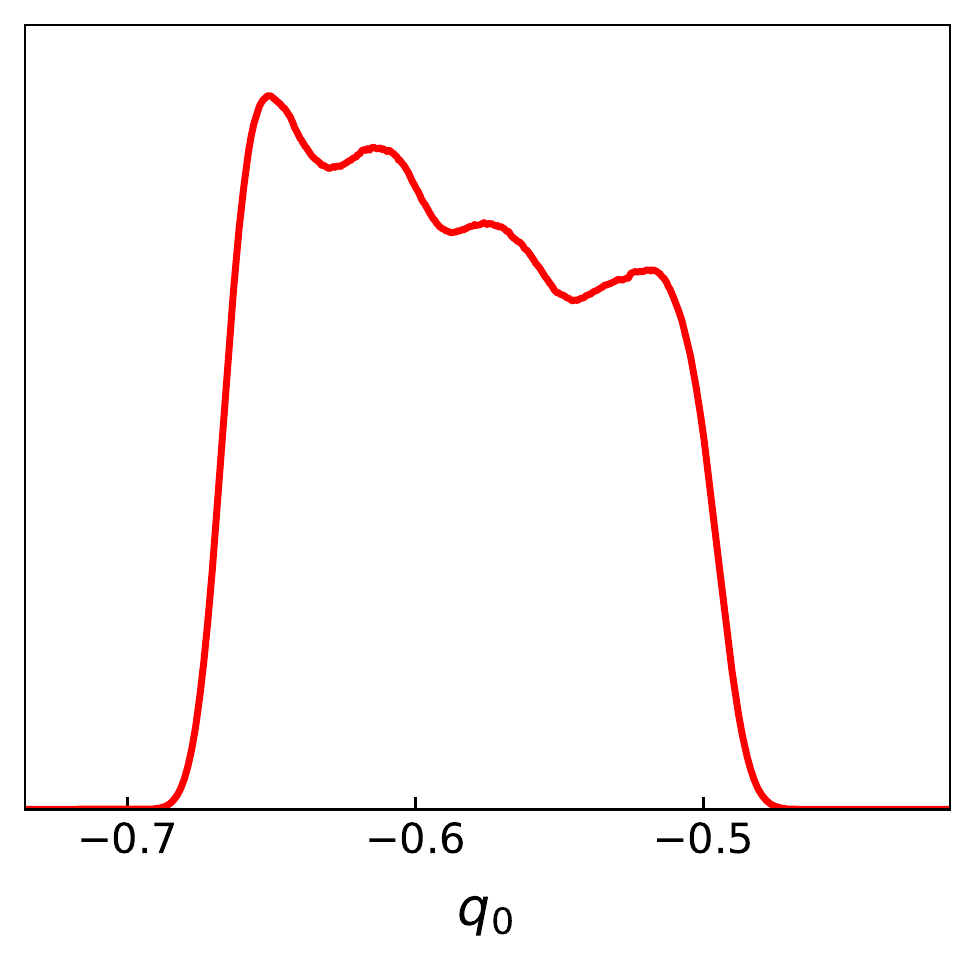}
\qquad
\includegraphics[width=8.5cm, height=7.5cm]{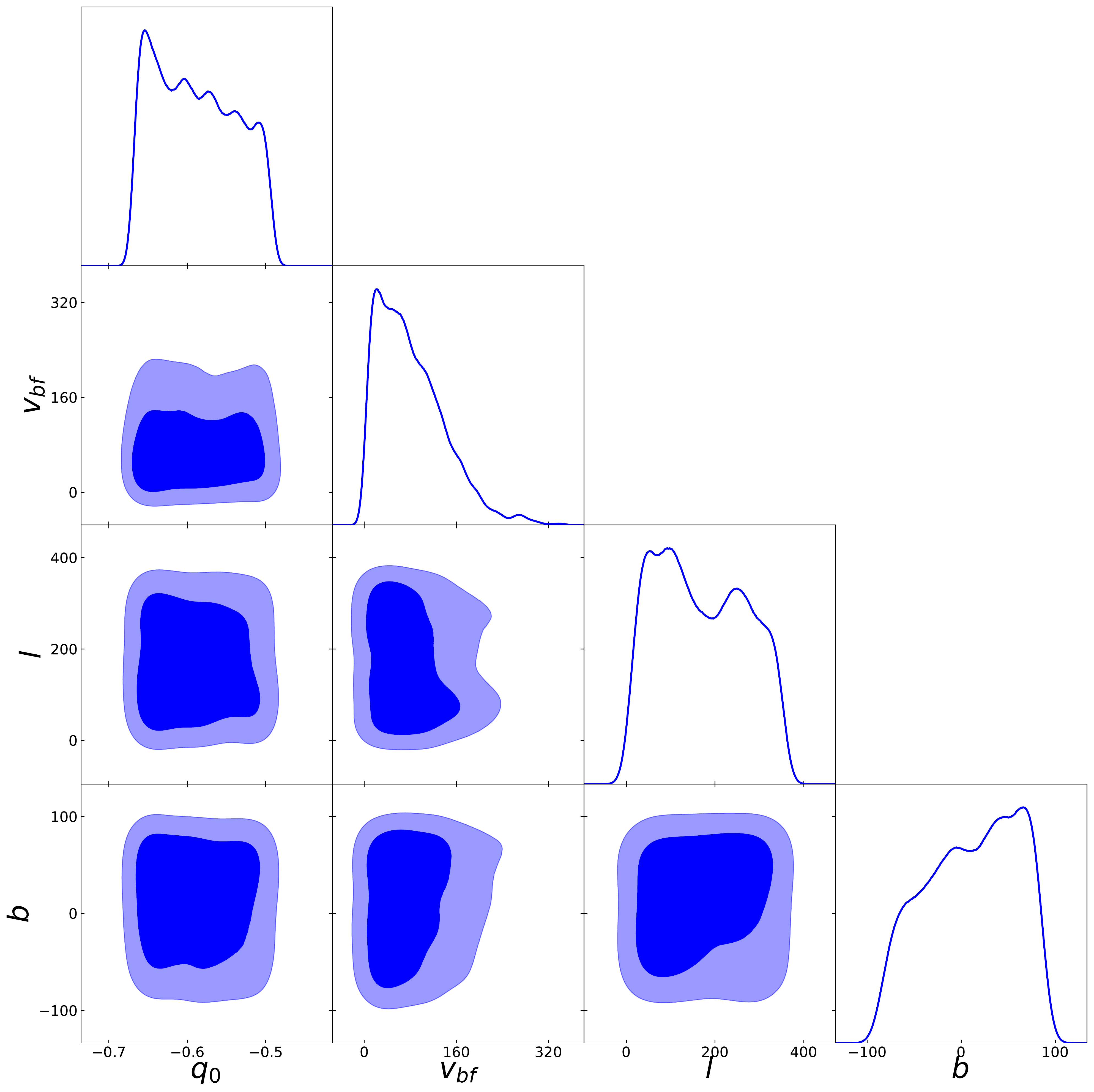}
\caption{The posteriors of the reference (left panel) and dipole (right panel) models, as obtained for the JLA compilation.}
\label{fig:post_JLA}
\end{figure}

\begin{figure*}[t]
\includegraphics[width=4.5cm, height=4.5cm]{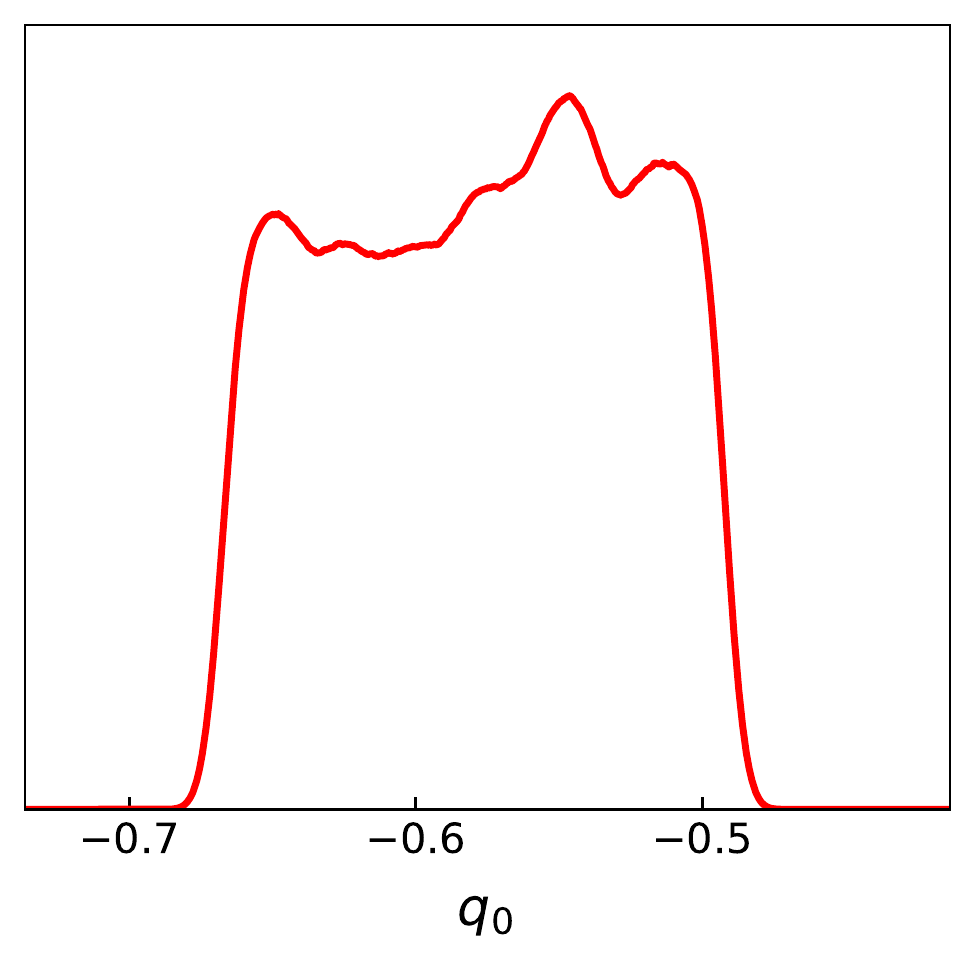}
\qquad
\includegraphics[width=8.5cm, height=7.5cm]{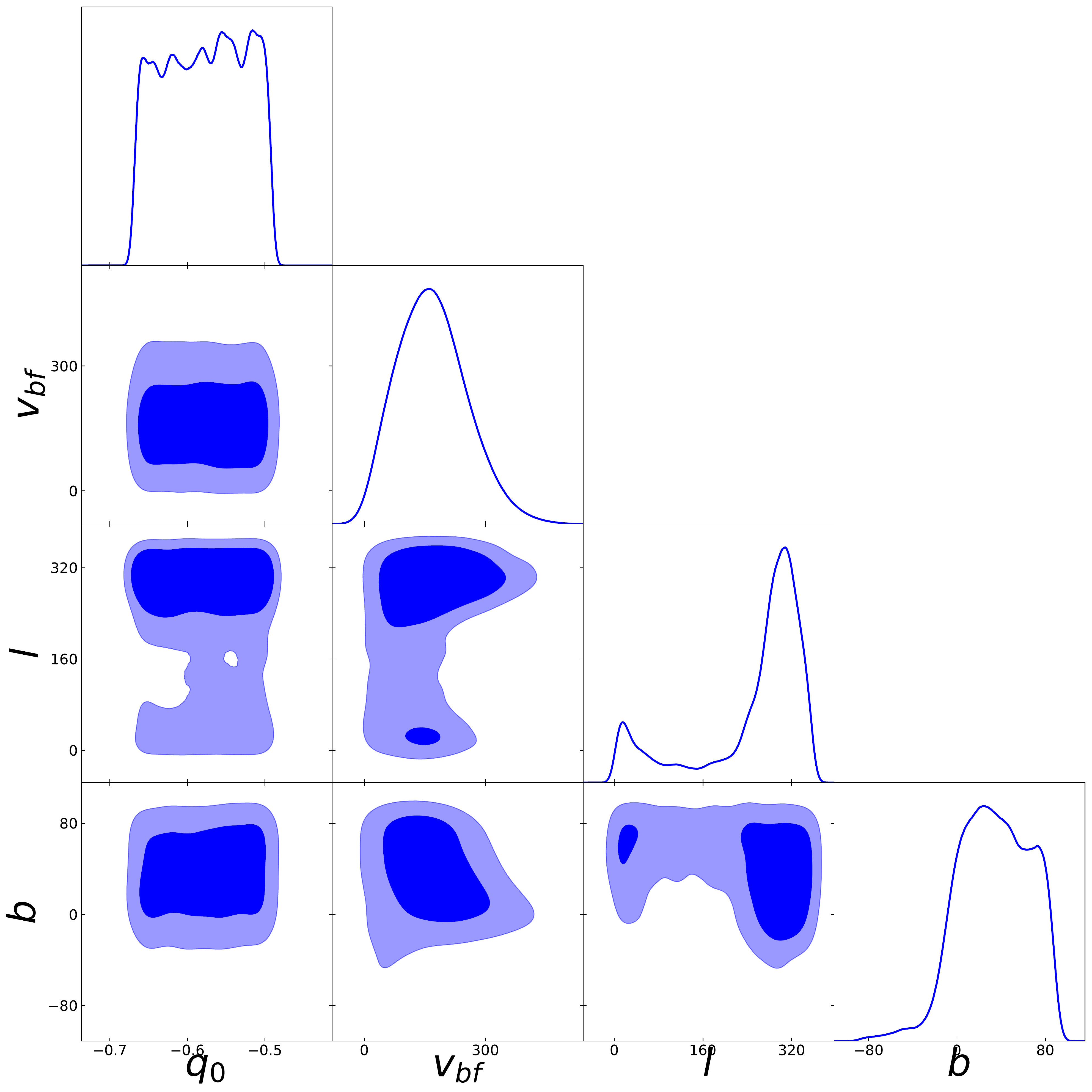}
\caption{Same as Fig.~\ref{fig:post_JLA} for the Union2.1 dataset.}
\label{fig:post_U21}
\end{figure*} 

Furthermore, we assume that both JLA and Union2.1 SN datasets follow a multivariate Gaussian likelihood, such as
\begin{eqnarray}\label{eq:likelihood}
\mathcal{L}(D| \Theta) \propto \exp{\left[ -\frac{\chi^2(D| \Theta)}{2} \right]} \;,
\end{eqnarray}
 whose $\chi^2$ reads
\begin{eqnarray}\label{eq:chi2}
\chi^2(D| \Theta) = \left[ \mat{\mu} - \mat{\mu} (\Theta\right)]^T C^{-1} \left[ \mat{\mu} - \mat{\mu} (\Theta\right)],
\end{eqnarray}
where $\mat{\mu}$ denotes a vector of the observed distance modulus, $\mat{\mu}(\Theta)$ are the theoretical values obtained from the models we shall test, and $C$ is the covariance matrix  which accounts for both statistic and systematic uncertainties for each sample.

In order to carry out the Bayesian analysis, we use the PyMultiNest \cite{buchner2014x}, a Python module based on the nested sampling (NS) algorithm MultiNest (\cite{feroz2008multimodal}, \cite{feroz2009multinest}, \cite{feroz2013importance}). Here, we make use of Importance Nested Sampling, since it can provide the Bayesian evidence with better accuracy than vanilla NS method according to~\cite{feroz2013importance}. Moreover, we assume flat priors on the set of cosmological parameters $\Theta$, as described in Table~\ref{priors}. Hence, we can determine how does the dipole model describe the observational data when compared to the reference one.

\begin{table}[t] 

\caption{The constraints on the parameters of reference model, denoted as Cosmography (CG), as well as the parameters characterizing the Dipole Model (DM). The last column shows the $95\%$ credible intervals.}

\renewcommand{\arraystretch}{1.5}
\renewcommand{\tabcolsep}{0.4cm}
\centering
\medskip
\label{table2}

	\begin{tabular}{cccccc}
		\hline
		\hline
		Dataset & Models & Parameters & Mean & Std. dev. & 95\% c.i.  \\
		\hline 

		&\textbf{CG} & $q_0$ & -0.587  & 0.052 & (-0.667, -0.497) \\ \\
		\textbf{\Large{JLA}}
		&& $q_0$ &  -0.589  &  0.051 &  (-0.666, -0.496) \\	
		&{\textbf{DM}} & $v_{bf}$ & 80.823 &  59.250 &   (3.705, 221.825) \\
		&& $l$ & 169.268  &  101.506   &(11.702, 348.162) \\
		&& $b$ & 11.630   & 49.062 &  (-80.856, 86.017) \\					
		\hline \hline
		

		&{\textbf{CG}} & $q_0$ & -0.578 & 0.051 & (-0.665, -0.495)   \\ \\
		\textbf{\Large{Union}}
		\textbf{\Large{2.1}} && $q_0$ &-0.577 & 0.052 & (-0.665, -0.494) \\
		&{\textbf{DM}} & $v_{bf}$ & 167.368 &   87.421 & (18.474, 352.066)  \\
		&& $l$ & 250.222  &  100.443 & (7.531, 353.462)   \\
		&& $b$ & 34.117 & 32.071 & (-28.728, 86.963)  \\ 	
		\hline
		\hline
	\end{tabular} 
\end{table}	
	

	\begin{table}[t]
		\caption{Bayesian evidence and Bayes' factors between the reference model and the dipole modulation one given the JLA and Union2.1 datasets.}
		\renewcommand{\arraystretch}{1.5}
		\centering
		\medskip
		\label{table3}
		\begin{tabular}{ccccc}
			\hline
			\hline
			Dataset & Models & ln$\mathcal{E}$ & ln$B$ & Evidence interpretation \\ 
			\hline
			& $CG$ & $-75.718 \pm 0.017$   & 0 &  ... \\ 
			\raisebox{2.ex}{\textbf{JLA}} & \textit{DM} & $-77.300 \pm 0.019$  & $1.582 \pm 0.026$ &  Weak (disfavored) \\
			& $CG$ & $-82.822 \pm 0.008$ & 0 &  ... \\ 
			\raisebox{2.ex}{\textbf{Union2.1}} & \textit{DM} & $-82.548 \pm 0.026$
			& $-0.275 \pm 0.027$ &  Inconclusive \\		
		\hline
		\hline
		\end{tabular} 
	\end{table}


\section{Results}

We show the mean estimates obtained from JLA for the parameters describing the reference and dipole model in the upper part of Table~\ref{table2}. Their corresponding posteriors are then presented in Fig.~\ref{fig:post_JLA}.  We note that the constraints on the cosmographic parameter $q_0$ do not appreciably change from one case to the other, and that there is just marginal evidence for a non-null bulk flow velocity from this dataset. Such a result is compatible with those reported in~\cite{huterer2015no}, where no significant evidence for a bulk flow using the same SN compilation was found. 

The lower part of the same Table~\ref{table2} presents the results obtained from the Union2.1 compilation. Again, the $q_0$ constraints are similar for both models, and the bounds imposed on the bulk flow velocity and direction are weakly restrictive. However, the mean bulk flow direction estimated from Union2.1 is compatible with previous analyses in the literature, such as the preferred direction reported by~\cite{colin2011probing}, which points towards $(l,b) = (316^{\circ},14^{\circ})$, as well as the results from~\cite{mathews2016detectability}, which obtained a bulk flow velocity located at the $(l,b) = (295^{\circ},10^{\circ})$ direction. 

Such a result is reflected in the model selection analysis, as presented in the upper part of Table~\ref{table3} for the JLA, and in the lower part for the Union2.1 dataset. We obtained weak evidence in favor of the reference model ($\ln{B} = 1.582 \pm 0.026$) for the former, while the results are inconclusive ($\ln{B} = -0.275 \pm 0.027$) for the latter\footnote{The error bars of the Bayesian factor and evidence correspond to the precision of the Multinest code on computing these quantities due to the sampling method it relies on.}. Although previous analysis reported that the absence of a dipole correction in $D_{L}$ is disfavored at $>2\sigma$ confidence level from an earlier low-redshift assemble of 44 SNe~\cite{bonvin2006dipole} adopting a maximum likelihood analysis, we obtained that the dipole model is not supported by a Bayesian model comparison analysis within the limitations of current SN compilations. 



\section{Discussion and conclusions}

In this work, we have performed a Bayesian comparison between two FLRW cosmological distance approaches, which differ from each other by a dipole modulation that arises from first order perturbation. This correction accounts for peculiar velocity due to local bulk motions, and therefore can provide constraints on the isotropy of the local Universe, i.e., at $z<0.1$. Because a FLRW Universe strongly relies upon the assumption of large-scale statistical isotropy and homogeneity, we need to assess whether they actually hold true in light of observational data. Any compelling evidence against such hypotheses would have profound implications for the standard paradigm of Cosmology. 

In our analysis, no further assumption on the dynamics of the Universe was made, as our description of the cosmological distances are purely given in terms of a Taylor expansion that gives rise to kinematic parameters, i.e., the cosmographic approach. We have truncated this expansion in redshift up to second order, since it accurately describes the luminosity distance at the scales we are probing, and thus compared it against a model which presents a correction in $D_{L}$ due to the nearby matter inhomogeneities~\cite{bonvin2006dipole} in the form of a dipole modulation. This comparison is carried out in a Bayesian framework for the first time, so that a strong evidence for a large anisotropy in the SN data would indicate a potential violation of the CP, at least of local origin.       

Adopting the two largest compilations of SNe currently available, namely JLA and Union2.1, in the low-z threshold ($z<0.1$), we have found weak evidence in favor of the reference model in the former, and inconclusive evidence for the latter, once both the statistical and systematic uncertainties of these samples are properly accounted for. This result is in agreement with the estimate from~\cite{huterer2015no} regarding the bulk flow constraints from JLA. Yet our Union2.1 result is compatible with previous analyses, we found no significant evidence in favor of the dipole modulation model in a Bayesian framework.     

Because blind tests of cosmological isotropy using SN data provided no statistical significance against the cosmic isotropy assumption as well~\cite{cooke2010does, antoniou2010searching, kalus2013constraints, Chang:2014jza, bengaly2015probing, Lin:2015rza, javanmardi2015probing, bengaly2016constraining, carvalho2016angular, ghodsi2016supernovae}, we conclude that there is no significant evidence for a dipolar anisotropy in the local Universe within the limits of present SN observations. A similar conclusion was drawn in~\cite{appleby2014testing}. As these authors used the luminosity function of $z<0.1$ galaxies rather than SNe, it provides an independent confirmation of our results, thus strengthening their significance. We expect this analysis to be highly improved once the next generation cosmological surveys, such as J-PAS~\cite{Benitez:2014ibt}, LSST~\cite{ivezic2008lsst}, Euclid~\cite{amendola2013cosmology}, and WFIRST~\cite{hounsell2017simulations} deliver much larger SN datasets with enhanced light-curve calibration, besides a stronger control of their potential systematics.


\acknowledgments

UA acknowledges financial support from Coordena\c{c}\~{a}o de Aperfei\c{c}oamento de Pessoal de N\'ivel Superior (CAPES). CAPB acknowledges the South African SKA Project, for financial support. JSA acknowledges support from CNPq (Grants no. 310790/2014-0 and 400471/2014-0) and FAPERJ (grant no. 204282). BS is supported by the DTI-PCI program of the Brazilian Ministry of Science, Technology, and Innovation (MCTI). Some of the analysis herein performed used the {\sc HEALPix} software package.  

\bibliography{ref_novo.bib}

\end{document}